\documentclass[a4paper,UKenglish,cleveref]{lipics-v2021}
\hideLIPIcs  

\title{DOMtutor: Automated Autograding for Logic in Computer Science}

\author{Tobias Meggendorfer}{Lancaster University Leipzig, Leipzig, Germany\and\url{https://tobias.meggendorfer.de} }{tobias@meggendorfer.de}{https://orcid.org/0000-0002-1712-2165}{}

\authorrunning{T.\ Meggendorfer}
\Copyright{Tobias Meggendorfer}


\category{Shared-Time Demo}

\supplement{\url{https://github.com/orgs/DOMtutor/}}

\nolinenumbers 

\EventEditors{Shriram Krishnamurthi and Thomas Zeume}
\EventNoEds{1}
\EventLongTitle{TEAL 2026: Tools for Educational Activities in Logic}
\EventShortTitle{TEAL 2026}
\EventAcronym{TEAL}
\EventYear{2026}
\EventDate{July 25, 2026}
\EventLocation{Lisbon, Portugal}
\EventLogo{}
\SeriesVolume{1}
\ArticleNo{1}

\usepackage{csquotes}
\usepackage{hyperref}

\begin{document}

\maketitle

\begin{abstract}
Teaching computer science at universities is often structured rather classically and theory oriented.
The former refers to \enquote{transmission}-style lectures accompanied by exercises which are submitted and graded manually, providing delayed feedback (if any).
The latter refers to exercises often posed at a conceptual level, requiring solution ideas to be sketched out on paper, but not put to the test in practice.
By its nature, this is particularly true for subjects relating to theoretical computer science, such as courses on propositional or first-order logic or automata theory.
Frameworks that automatically execute and evaluate code (also called \emph{autograders}) are sometimes used to augment teaching.
They provide (near) instant feedback and hands-on experience, prompting reflective analysis.
However, their use usually is reserved for programming / practically oriented courses.
We propose to (i)~use autograders also (and especially) for theoretical courses and (ii)~use the established \texttt{DOMjudge} system, which is used, among others, for the International Collegiate Programming Contest.
\end{abstract}

\section{Introduction}
University-level computer science education historically relied on non-personal, lecture-centred formats, and transmission-based teaching \cite{kay1998large}.
This is exacerbated by constantly growing student numbers, which make individualized support unrealistic \cite{hornsby2014massification}.
In turn, this often leads to streamlined, simple-to-grade exercises and exams, to keep the workload of teaching staff manageable \cite{whisenhunt2019strategies}.
In particular, formative exercises (i.e.\ ones not relevant for the grade) are usually not given any feedback or only a single instance of delayed feedback (submit the exercise, wait days or weeks for grading, receive the feedback).
While often necessitated by a practical lack of (human) resources, this encourages shallow learning \cite{biggs1996enhancing}, leaving little space for reflection and exploration (i.e.\ \enquote{why did this go wrong?} or \enquote{what happens if I try this?}), both of which significantly improve student performance \cite{freeman2014active}.
Overall, this often leaves students with a feeling of mass processing, which in turn lowers engagement.

One method to (partially) address these issues, in particular by providing immediate and individualized feedback, is given by \emph{autograder} (or automated assessment system) platforms.
Broadly interpreted, such tools allow students to submit solution attempts, which are automatically graded, and feedback is given immediately, possibly also allowing to try again.
This encompasses a wide variety of tools, even including simple online quizzes as provided by e.g.\ Moodle.
A particularly appealing sub-category (which is often used synonymously) are autograders for programming, where students submit source code, which can then be tested against unit tests, evaluated for code-style, etc.
Appropriately, these systems are comparatively popular in coding-oriented contexts, such as introduction courses to programming, as they naturally align with the learning outcomes of such modules \cite{messer2024automated}.
There exists a broad variety of autograders, operating on different \enquote{levels} -- some focus on input-output behaviour, treating the program as a black box, others tie in through language specific unit tests or code analysis.
Similarly, the feedback can be binary (\enquote{correct} or \enquote{incorrect}), more fine grained (\enquote{4 of 20 test cases failed}) or with additional feedback (\enquote{incorrect because of the following corner case}).

Unfortunately, autograders also come with several hurdles, which make them (seemingly) unappealing in many contexts.
Firstly, they naturally require more infrastructure than simple pen-and-paper exercises, such as installing and maintaining the tool, providing protected and privacy-conformant access for students, high availability, reliable backup solutions, and so on.
Secondly, designing exercises is technically more challenging than writing a paragraph of text (requiring, e.g., the proper definition of test cases or sample solutions).
Relatedly, being able to automatically grade necessitates an unambiguous problem statement, a computable testing procedure, and a stable implementation thereof.
While manually graded exercises allow for some leeway (\enquote{this is not the intended solution approach but correct nonetheless}), automatically gradable exercises require significantly more care from the instructor, also to avoid student dissatisfaction due to misleading feedback, such as an actually correct solution being marked as incorrect.
Finally, due to their branding (\enquote{coding autograder} and the like), such tools are often only considered for programming-related courses.

This contribution comprises two parts:
First and foremost, we postulate that coding autograders can be very appealing and useful in many more contexts, even in theoretical courses relating to, e.g., automata theory or linear algebra.
This is discussed in \cref{sec:theory}.
Secondly, by using the long-existing and maintained \texttt{DOMjudge}\footnote{\url{https://www.domjudge.org/}} platform together with our herein presented tool \texttt{DOMtutor}, a lot of the practical overhead is taken care of, allowing instructors to focus on high-quality teaching and use autograders without spending days on setup.
This is discussed in \cref{sec:domjudge}.
Finally, \cref{sec:demo} summarizes the planned content of the shared demo.

\section{Programming for Theory} \label{sec:theory}
As mentioned, autograders are hardly considered for theory, as they are naturally associated with programming problems.
However, we argue that they can play an invaluable role even in traditionally purely theoretic subjects, such as logic, linear algebra, or automata theory.
The driving idea of using autograders is the importance and effectiveness of broadly available instant feedback for learning.
By providing an online platform that can directly judge an attempt, students can learn and try things on their own pace, and engage with the topic at hand in a reflective manner \cite{sauerwein2023towards}.
Moreover, precise, unambiguous grading can enforce also considering corner cases, instead of just sketching the main idea without fully fleshing it out, as often is the case in pen-and-paper exercises.
As also confirmed by numerous instances of student feedback in the author's lectures, being required to fully \enquote{think through} the problem and explicitly considering potential corner cases notably deepens understanding \cite{barczak2023automated}.

Observe that the aforementioned points are not specifically tied to classical programming tasks, but ubiquitous to learning in general, suggesting that autograders are applicable in a wide range of topics.
To support this claim, we outline several examples which have been successfully used in theory lectures, receiving praise from the students.
For now, we leave the exact kind of autograder open, as these ideas can be implemented in a broad variety of tools (with differing levels of complexity).
One possible instance is discussed later.

\paragraph*{Automata Theory}
Our first example relates to automata theory.
In an introductory course, there is practically no way around learning deterministic and non-deterministic finite automata and numerous procedures related to them, e.g.\ word membership, complementation, determinization, etc.
While all these concepts are often taught on a theory level and executed by hand on small examples, they also lend themselves naturally to implementation, even more so as one often is a sub-routine to the other, which allows for further scaffolding.
As such, when first introducing automata, one can provide students with an easy to understand skeleton implementation of automata and ask them to implement the word inclusion check.
This builds basic understanding of the algorithmic foundations.
Then, in the next step, determinization can be implemented, and, as additional exercise, complementation of the language.
Notably, checking the correctness of the computed result can easily be verified algorithmically.
Already at such a \enquote{straightforward} level, correct algorithms need to handle corner cases.
For example, once $\varepsilon$-NFA are introduced, it is easy to miss handling $\varepsilon$-cycles in their determinization procedure.
This is a subtle point that often goes unnoticed in manual exercises, as there the visited states are tracked implicitly.
Here, an instructor can add test cases (likely hidden from the students) that specifically tests for correct handling of such cycles.

\paragraph*{Linear Algebra}
Clear problems in teaching linear algebra are, for example, matrix-vector multiplication, computation of the determinant, matrix inversions, determining eigenvectors, or even decomposition.
First tasks could be restricted to integer matrices of a fixed size, to simplify implementation, yet still allow to explore the basic concepts of the topic.
Corner cases could include correct handling of repeated eigenvalues.
When linear algebra is being taught to computer science students, it however lends itself well to also demonstrate what happens if theory meets practice.
For example, one can then move to matrices given as floating point numbers and demonstrate how quickly naive implementations using IEEE754 floating point numbers can fail, how significant the overhead of using arbitrary precision arithmetic is, or how to detect numerical issues and ill-conditioned problems.
Here, concrete implementations with concrete inputs allow to tangibly experience how such issues materialize in practice, instead of just considering abstract criteria and proving their implications.

\paragraph*{Logic}
Propositional logic allows for many natural tasks, too.
For example, checking whether a given valuation satisfies a given CNF, converting a general formula to an equisatisfiable CNF, computing the negation of a CNF formula in CNF, checking satisfiability or validity, and so on.
However, we want to use this as a different example, where motivated students can go beyond the core content of a lecture.
For example, when learning about first order logic in general, students can be given a longer-running task to implement quantifier elimination for Presburger arithmetic.
As a clearly marked extra challenge, this allows them to further their understanding of the topic.

\subsection*{Discussion}
As described above, even traditionally theoretical topics can be well supported by practical, programming-oriented tasks.
Notably, since this is discussed in the context of computer science curricula, the students can be expected to already have a basic understanding of at least one programming language (with the added benefit of recalling and further sharpening programming skills).
Depending on the progression within the curriculum and the \enquote{difficulty} of the languages taught before (e.g.\ C vs.\ Python), skeletons of different depth can be provided to students, taking care of e.g.\ appropriate representation of the underlying data structures.
In terms of suitable tasks, any decidable problem can be posed as a programming exercise, and even undecidable one by focusing on fragments or giving additional guarantees (\enquote{if a solution exists it has size at most X}).
However, some are more manageable than others.
For example, implementing a DPLL-based SAT solver is substantially easier than a sound and complete decision procedure for the existential theory of the reals.

We also underline that we do not suggest to replace in-person workshops by such exercises.
They rather comprise meaningful and scalable addition to the overall teaching framework, providing yet another learning modality.
In particular, they work well in tandem with in-person workshops, where the concepts from the lecture can be recalled, and then later put to practice in such implementation exercises.

Together, we strongly believe that programming tasks are well suited to augment existing teaching structures by providing a new, tangible perspective on potentially theoretical topics and, backed up by anecdotal evidence, advocate for their use where feasible.

\section{Theory to Practice: DOMjudge and DOMtutor} \label{sec:domjudge}
The previous section discussed how autograders in general can be used to support learning, even in theoretically oriented courses.
There exist a wide variety of autograders, and many institutions (or even individual teaching staff) additionally develop their own, tailored tools.
Employing an \enquote{in-house} build however comes with some friction:
A lot of technical details (e.g.\ process sandboxing) need to be solved, maintenance needs to be ensured, etc., often leading to ad-hoc and undocumented implementation.
As such, these tools often are \enquote{attached} to individual staff rather than lectures or institutions, which in turn reduces the chance of re-use.
Here, we want to present an established autograder together with our (hopefully) flexible management script set, which together are well suited to support all the above problems (and many more).
We intentionally omit describing the technical details, as they are out of scope.

\paragraph*{Autograder Environment}
We believe that \texttt{DOMjudge} stands out as an autograder platform due to several reasons.
Originally, it has been designed for student programming contests, most notably the International Collegiate Programming Contest (ICPC) and its regional variants.
It has been in continuous use for such contests since over two decades and is still being actively developed.
Since autograders execute untrusted code, they require tight sandboxing, which comes with many subtle challenges -- these are all solved by \texttt{DOMjudge}.
By its focus on programming contests, it is designed for programming language independence from the ground up, which actually makes it suitable even for non-programming tasks (with dedicated support also currently being developed further).
It relies on the separately developed Kattis problem format\footnote{\url{https://www.kattis.com/problem-package-format/}}, which similarly is designed with openness in mind.
Moreover, it directly supports deployment via Docker and uses a standard database, making setup and backup straightforward.
Altogether, \texttt{DOMjudge} is one of the most mature, versatile, and broadly used graders, making it a good candidate for employing automated exercises in teaching.
While it also has some limitations, we believe that it can cover a very broad spectrum of teaching use-cases.
And indeed, several institutions also use \texttt{DOMjudge} for their teaching.

\paragraph*{Autograder Management}
While \texttt{DOMjudge} is well suited for handling the actual submission and evaluation, its setup capabilities and post-processing are somewhat limited.
To this end, we developed \texttt{DOMtutor}, a large suite of Python scripts which automate managing \texttt{DOMjudge} instances.
For example, \texttt{DOMtutor} reduces synchronizing the list of users with a Moodle instance to a push-button task.
It allows to export results of an exercise, generate various statistics, and simplify assigning a final grade based on one or multiple submissions.
In general, it provides a programmatic interface to all relevant parts of \texttt{DOMjudge}.
We also develop it further, partly through student projects, to, for example, provide an AI-driven \enquote{problem studio}, which simplifies problem design by letting an LLM draft a concrete problem (see also \cite{alkafaween2025automating}), or an implementation of plagiarism detection / pre-filtering (using, among others, \texttt{copydetect}\footnote{\url{https://github.com/blingenf/copydetect}} and \texttt{RapidFuzz}\footnote{\url{https://github.com/rapidfuzz/RapidFuzz}}).

Overall, \texttt{DOMtutor} is (i)~a python library to access and manipulate \texttt{DOMjudge}, (ii)~an (opinionated) implementation of a course workflow that integrates well into standard lecture-workshop based teaching, and (iii)~a broad set of reusable utilities such as a \enquote{fuzzer} (which generates small counterexamples for failing student code) and a simple bot that monitors the state of the \texttt{DOMjudge} instance.
It significantly reduces management overhead and allows for high degree of customization by simply adapting high-level scripts.
It has been successfully used by the author for over ten instances of five different lectures.

\subsection*{Discussion}
While there are many possible platforms out there, we advocate the use of \texttt{DOMjudge} due to its maturity and (comparable) flexibility.
Flexibility of course comes with slight overhead to \enquote{massage} a concrete case into the general format, however we believe that a large set of problems can be phrased in the Kattis format with comparatively limited effort.
\texttt{DOMjudge} is paired with \texttt{DOMtutor}, which simplifies management of larger and long-running setups as well as integration with lectures.
Together, they solve a wide spectrum of \enquote{annoying} technical obstacles, which would typically prevent use of autograders.
We expect that (with some support) novice users can set up and configure the entire system to the point that students can submit attempts to a custom problem within a few hours.
However, we underline that we do not claim that this pairing is without any alternatives.

\section{Shared-Demo} \label{sec:demo}
As mentioned, our main interest is to propagate the idea of employing programming tasks even in seemingly theoretical contexts.
Firstly, this reduces workload of instructors especially for large courses.
Secondly, this potentially increases student engagement through providing a new avenue towards the taught content, as a \enquote{hands-on} experience often makes theoretical topics more tangible.
To pave the road for this use, we also propose the use of our \texttt{DOMtutor} script set, which eases practical management of such tasks.
Our proposed demo is twofold.
Firstly, during the actual demo session, we want to discuss the participants' individual use cases and explore how \texttt{DOMjudge} and \texttt{DOMtutor} can be tied into this particular setting.
This also will help drive the further development of \texttt{DOMtutor}.
We also want to show the capabilities of both tools and given an example of both the student's and instructor's user experience.
Secondly, we will set up a \texttt{DOMjudge} instance and create several theory-related challenges, similar to the ones mentioned above, that run over the entire duration of FLoC, together with language skeletons.
This allows participants to experience the tool first hand (and apply some potentially long unused undergraduate knowledge).

\bibliography{main}
\end{document}